\documentclass[11pt,twoside,letterpaper]{article}
\usepackage{amssymb,amsfonts,amsbsy}
\newcommand{\be}{\begin{equation}}
\newcommand{\ee}{\end{equation}}
\newcommand{\ba}{\hspace*{-5pt}\begin{array}}
\newcommand{\ea}{\end{array}}
\newcommand{\p}{\partial}

\def\Im{\mathop{\rm Im}\nolimits}
\DeclareMathAlphabet{\bi}{OML}{cmm}{b}{it}

\def\rank{\mathop{\rm rank}\nolimits}

\def\Im{\mathop{\rm Im}\nolimits}

\newcommand{\bu}{\bi{u}}

\def\Im{\mathop{\rm Im}\nolimits}

\newtheorem{lem}{Lemma}
\newtheorem{cor}{Corollary}
\newtheorem{prop}{Proposition}

\title
{\vspace*{-10mm}\protect\bf A simple way
of making a Hamiltonian system\\ into a bi-Hamiltonian one}

\begin{document}
\author{A. Sergyeyev\thanks{On leave from Silesian University in Opava,
Mathematical Institute, Na Ryb\-n\'\i{}\v cku 1,
746~01 Opava, Czech Republic. E-mail: {\tt Artur.Sergyeyev@math.slu.cz}}\\
The Jacob Blaustein Institute for Desert Research,\\
Ben-Gurion University of the Negev, Sede Boqer Campus,\\ 84990 Israel
}
\date{}
\maketitle
\begin{abstract}
Given a Poisson structure (or, equivalently, a Hamiltonian
operator) $P$, we show that its Lie derivative $L_{\tau}(P)$ along
a vector field $\tau$ defines another Poisson structure, which is
automatically compatible with $P$, if and only if
$[L_{\tau}^2(P),P]=0$, where $[\cdot,\cdot]$ is the Schouten
bracket. We further prove that if $\dim\ker P\leq 1$ and $P$ is of
locally constant rank, then all Poisson structures compatible with
a given Poisson structure $P$ on a finite-dimensional manifold $M$
are locally of the form $L_{\tau}(P)$, where $\tau$ is a local
vector field such that $L_{\tau}^2(P)=L_{\tilde\tau}(P)$ for some
other local vector field $\tilde\tau$. This leads to a remarkably
simple construction of bi-Hamiltonian dynamical systems.
We also present a generalization of these results to the
infinite-dimensional case. In particular, we provide
a new description for pencils of compatible local Hamiltonian operators of
Dubrovin--Novikov type and associated bi-Hamiltonian systems of
hydrodynamic type.
\looseness=-1

\looseness=-1

\end{abstract}
%
\section{Introduction}
The ingenious discovery of Magri \cite{magri} (cf.\ also \cite{gd, mamo, ff}
and the surveys in \cite{dor, olv_eng2, bl})
that integrable Hamiltonian systems usually prove to be bi-Hamiltonian,
and vice versa, leads us to the following fundamental problem:
given a (likely to be integrable) dynamical system which is
Hamiltonian with respect to a Poisson structure $P$, how to find
another Poisson structure $\tilde P$ compatible with $P$ and such
that our system is Hamiltonian with respect to $\tilde P$ as well.
This, along with
the related problem of classification of compatible
Poisson structures, is nowadays a subject of intense research,
see e.g.\ \cite{magri}--\cite{pet} and references therein.
\looseness=-1

For the finite-dimensional dynamical systems the results of
Lichnerowicz \cite{li} imply that if $\dim\ker P\leq 1$ and $P$ is
of locally constant rank (here and below we assume that the vicinities
where $\rank P=\mathrm{const}$ are of the same dimension as the
underlying manifold), then {\em all} Poisson structures compatible
with $P$ are locally
(and, under certain extra conditions, globally)
of the form $L_{\tau}(P)$, i.e.,
they can be written as Lie derivatives of $P$ along
suitable local vector fields $\tau$. Oevel \cite{oevth}
and, independently,
Dorfman \cite{dor} showed that for invertible $P$
this holds in the infinite-dimensional case as well.
Oevel \cite{oevth, oev} also pointed out that the $\tau$'s
in question often prove
to be {\em master symmetries} in the sense of \cite{ffms}.
\looseness=-1

In general $L_{\tau}(P)$ is not
a Poisson structure
even if so is $P$, and it is our goal here
to provide a simple description
of the `eligible' $\tau$'s, for which
$L_{\tau}(P)$ {\em is} Poisson. Namely,
see Proposition~\ref{cpb2vfloc} below,
we prove that if $\dim\ker P\leq 1$ and $P$ is of locally constant rank, then
{\em all} Poisson structures compatible with $P$ are locally of
the form $L_{\tau}(P)$, where $\tau$ is a local vector field such
that $L_{\tau}^2(P)=L_{\tilde\tau}(P)$ holds locally for some
other local vector field $\tilde\tau$. For invertible $P$ this
holds even in the infinite-dimensional case, and under certain
conditions $\tau$ and $\tilde\tau$ are global, see
Proposition~\ref{cpb2vf} and Section \ref{infdim} below for
details. This improves earlier results of Petalidou, who found a
criterion for $L_{\tau}(P)$ to be a Poisson structure under the
assumption that $P$ is nondegenerate, see Proposition~3.1 of
\cite{pet}, and also \cite{peth}.
Moreover, Propositions~ \ref{cpb2vf} and \ref{cpb2vfloc}
yield a criterion for existence of a regular or weakly regular Poisson pencil
such that a given dynamical system is locally bi-Ha\-mil\-tonian
with respect to this pencil, see Remark 1 and Corollary~\ref{intloc}.

On the other hand, 
if both $P$ and $L_{\tau}(P)$ {\em are} Poisson structures, then
they are {\em automatically} compatible \cite{li, oevth, mma,
dor}, no matter whether $P$ is (non)degenerate, so it is natural
to ask when $L_{\tau}(P)$ is a Poisson structure if so is $P$.
We show that this is the case if and only if $[L_{\tau}^2(P),P]=0$,
where $[\cdot,\cdot]$ is the Schouten bracket,
see Proposition~\ref{biham} below for details.
Note that this result first appeared
in the earlier version \cite{bih-prepr}
of the present paper and
was recently rediscovered by Chavchanidze
\cite{chav} in the context of non-Noether symmetries.\looseness=-2

By Corollary~\ref{inf_st}
if there exist a (local) vector field $\tilde\tau$ and
a constant $\alpha$ such that $L_{\tau}^2(P)=L_{\tilde\tau}(P)+\alpha P$,
then $[L_{\tau}^2(P),P]=0$, and thus $L_{\tau}(P)$ is a Poisson structure.
For $\alpha=0$ and $\tilde\tau=0$ we recover the result of
Smirnov \cite{smi}, cf.\ also formulae (2.8) in
Magri \cite{magri2} and the papers \cite{cas, fal}; in this
case
$P$ and
$L_{\tau}(P)$ form the so-called {\it exact bi-Hamiltonian
structure}, and $\tau$ is called a {\em Liouville vector
field} \cite{cas, fal}. On the other hand, given
two compatible Poisson structures
$P$ and $\tilde P$ that share a common
scaling and assuming that one of them, $P$, is nondegenerate,
we can readily construct
$\tau$ and $\tilde\tau$ such that the second bivector of the pair, $\tilde P$,
can be written as $L_{\tau}(P)$
and $L_{\tau}^2(P)=L_{\tilde\tau}(P)$,
see Proposition~\ref{cpb}
and Corollary~\ref{cpbcor} below.
\looseness=-1

As an application of our results, in Section \ref{infdim}
we give a new description of compatible
Poisson structures of
Dubrovin--Novikov \cite{dn,dn84} type and of associated
bi-Hamiltonian systems of hydrodynamic type.
\looseness=-1

\section{Basic definitions}
Let $M$ be a smooth
finite-dimensional manifold.
Below we assume all
objects to be smooth enough 
for all the required
derivatives to exist.
\looseness=-1

Recall that a {\em bivector} on $M$ is a skew-symmetric
contravariant tensor field of rank two. To any bivector $B$ there
corresponds, in a natural way, a skew-symmetric linear operator
(which for the sake of simplicity will be denoted by the same letter)
$B:\wedge^1 M \rightarrow TM$. A bivector $B$ is
called {\em nondegenerate}, and the associated operator
$B$ is called {\em invertible},
if 
$\ker B=\{\chi\in\wedge^1 M: B\chi=0\}$ is
exhausted by $\chi=0$. \looseness=-1

The Schouten bracket $[H,K]$ of two bivectors $H$ and $K$
is given by the formula
(see e.g.\ Section 2.8 of \cite{dor})
\looseness=-1
$$
[H,K](\xi_1,\xi_2,\xi_3)=\langle H L_{K \xi_1}(\xi_2),\xi_3\rangle
+\langle K L_{H \xi_1}(\xi_2),\xi_3 \rangle + \mbox{cycle}(1,2,3),
$$
where $\xi_1,\xi_2,\xi_3\in\wedge^1 M$,
$L_X$ stands for the Lie derivative along a vector field $X$, and
$\langle\cdot,\cdot\rangle$ denotes the natural pairing of vector
fields and one-forms on $M$.

Thus, $[H,K]$ is an antisymmetric contravariant tensor of rank three, i.e., a trivector,
and its components in local coordinates read \cite{dor}
$$
[H,K]^{ijk}=-\sum\limits_{m=1}^{\dim M}\left(K^{mk} {\p H^{ij}\over\p x^m}
+H^{mk} {\p K^{ij}\over\p x^m}+\mathrm{cycle}(i,j,k)\right).
$$

It is well known \cite{mma, dor} that
\be
\label{sym} [H,K]=[K,H],
\ee
and for any vector field $\tau$ on $M$
\be
\label{sch_lie}L_{\tau}([H,K])=[L_{\tau}(H),K]
+[H,L_{\tau}(K)].
\ee
Recall that in local coordinates the Lie derivative of a bivector $P$
along a vector field $\tau$ reads
$$
(L_{\tau}(P))^{ij}=\sum\limits_{k=1}^{\dim M}\left(\tau^k{\p P^{ij}\over\p x^k}
-P^{kj}{\p\tau^i\over\p x^k} -P^{ik}{\p\tau^j\over\p x^k}\right).
$$

If $[P,P]=0$, then a bivector $P$ on $M$ is called a {\em Poisson
bivector} or, if it is perceived as an operator $P: \wedge^1 M
\rightarrow TM$, a {\em Hamiltonian} \cite{dor} or {\em implectic}
\cite{ff} operator. The associated Poisson bracket reads
$\{f,g\}_{P}=\langle d f, P d g\rangle$, where $f$ and $g$ are
smooth functions on $M$, see e.g. \cite{dor, bl}. A pair $(M, P)$,
where $P$ is a Poisson bivector on $M$, is called a {\em Poisson
manifold} \cite{li}.

Two Poisson bivectors $P_0$ and $P_1$ (or the associated
Hamiltonian operators) are said \cite{magri, gd, mamo} to be {\em
compatible} (or to form a  {\em Hamiltonian pair}), if any linear
combination of $P_0$ and $P_1$ is again a Poisson bivector. It is
well known, see e.g.\ \cite{magri, gd, mamo, dor}, that $P_0$ and
$P_1$ are compatible if and only if
$[P_0,P_1]=0$.

\section{When $L_{\tau}(P)$ is a Poisson bivector?}\label{whenpb}
\begin{prop}
\label{biham}
Let $P$ be a Poisson bivector and $\tau$ be a vector field on $M$.
Then $L_{\tau}(P)$ is a Poisson bivector, which is automatically
compatible with $P$, if and only if \be\label{comp} [L_{\tau}^2
(P), P]=0. \ee \looseness=-1
\end{prop}
%
{\em Proof.}
Writing out the identity $L_{\tau}([P,P])=0$
with usage of (\ref{sym}) and (\ref{sch_lie})
yields (cf.\ Proposition~7.8 from \cite{dor})\looseness=-1
\be\label{comp3}
\quad [P,P]=0\Rightarrow [L_{\tau}(P),P]=0. \ee

Next, using (\ref{sch_lie}) and 
(\ref{comp3}), we can rewrite 
the identity
$L_{\tau}^2([P,P])=0$ as $[L_{\tau}^2 (P), P]
+[L_{\tau}(P),L_{\tau}(P)]=0$. As $L_{\tau}(P)$ is a Poisson
bivector if and only if $[L_{\tau}(P),L_{\tau}(P)]=0$,
the result immediately follows. $\square$


For instance, let $M=\mathbb{R}^{2m+1}$
with 
(global) coordinates $x^i$ and
$h$ be
a smooth function on $M$.
Taking for $P$ the canonical Poisson structure
of maximal rank on $M$ and setting
$\tilde P=L_{\tau}(P)$
for $\tau=(-(x^1)^2/2, \dots,\allowbreak -(x^{m})^2/2, 0,
\dots,\allowbreak 0, -h/2)^{T}$, where
the superscript $T$ stands for the transposed matrix, we have
$$ P=\left(\begin{array}{rrr} 0 & I  & 0\\ -I & 0
& 0\\ 0 & 0 & 0\\
\end{array}\right), \quad
\tilde P=\left(\begin{array}{ccc} 0 & \Lambda  & X^T\\ -\Lambda
& 0 & -Y^T\\ -X & Y  & 0\\
\end{array}\right).
$$
Here $I$ is the $m\times m$ unit matrix, $\Lambda=\mathop{\rm
diag}\nolimits (x^1,\dots,x^m)$ is an $m\times m$ diagonal matrix,
$X=(\p h/\p x^{m+1},\dots,\allowbreak \p h/\p x^{2m})$ and $Y=(\p h/\p
x^{1},\dots,\allowbreak \p h/\p x^{m})$ are $m$-component columns.

For $h=\sum_{i=1}^{m}f_i(x^i, x^{i+m})/\Delta_i-
x^{2m+1}\sum_{i=1}^{m}x^i$, where $f_i$ are arbitrary
smooth functions of their arguments, and
$\Delta_i=\prod_{j=1,j\neq i}^m (x^i-x^j)$, this construction
yields a pair of Poisson structures
arising in the theory of the so-called one-Casimir chains
\cite{bl,bl1,bl2}.
It is straightforward to verify that in this case $\tilde P$ indeed is
a Poisson structure by virtue of~(\ref{comp}), and
$P$ and $\tilde P$ are compatible by (\ref{comp3}).
The examples of integrable systems associated with this pair can be
found in \cite{bl,bl1,bl2}.

Remarkably, (\ref{comp}) is often easier to verify than
$[L_{\tau}(P), L_{\tau}(P)]=0$. In particular, using (\ref{comp3})
and bilinearity of the Schouten bracket readily yields the
following result.
\looseness=-1
\begin{cor}\label{inf_st}
Let $P$ be a Poisson bivector and there exist vector fields $\tau$
and $\tilde\tau$ and a constant $\alpha$ such that
\be\label{comp2} L_{\tau}^2 (P)=L_{\tilde\tau}(P)+\alpha P. \ee
%
Then $L_{\tau}(P)$ is a Poisson bivector, which is automatically
compatible with $P$. \looseness=-1
\end{cor}

If we set $\alpha=\tilde\tau=0$ in (\ref{comp2}), then we
recover Proposition 4.1 of Smirnov \cite{smi} (cf.\ also
Section 2 of \cite{fal}). Note that for $P$ being a Poisson
structure of Dubrovin--Novikov type
\cite{dn,dn84} the condition (\ref{comp2}) with
$\alpha=\tilde\tau=0$ was also studied by Fordy and Mokhov
\cite{fm}. \looseness=-1

For an example of somewhat different kind, let $M=\mathbb{R}^3$ with the
(global) coordinates $x,y,z$ and
$\tau=(-x-x^3/3, 0, z x^2-x^2 y^2)^{T}$.
%
Take for $P$ the canonical Poisson structure on $\mathbb{R}^3$.
Then we have
\be\label{blrwst}
\hspace*{-2mm}
P=\left(\begin{array}{@{}ccc@{}} 0  & 1 & 0\\ -1 & 0  & 0\\ 0  & 0
& 0
\end{array}\right), 
L_{\tau}(P)=\left(\begin{array}{@{}ccc@{}} 0 & 1+x^2  & 2x^2 y\\
-1-x^2 & 0  & 2xz-2xy^2\\ -2x^2 y & -2xz+2xy^2 & 0
\end{array}\right).
\ee In turn, $L_{\tau}^2(P)=L_{\tilde\tau}(P)$, where
$\tilde\tau=(-x^5/15-x, 0, (z x^3-x^4 y^2)/3)^{T}$, so
$L_{\tau}(P)$ is a Poisson bivector by
Corollary~\ref{inf_st}.
This pair of compatible Poisson bivectors was found in \cite{blrw}
along with
a related completely integrable bi-Hamiltonian system.
\looseness=-1

Now assume that we have two vector fields $\tau_1$ and $\tau_2$
meeting the requirements of Proposition~\ref{biham}.
When are the Poisson bivectors $P_1=L_{\tau_1}(P)$ and $P_2=L_{\tau_2}(P)$
compatible?

\begin{cor}\label{2t}
Given a Poisson bivector $P$ and two vector fields $\tau_1,\tau_2$
meeting the requirements of Proposition~\ref{biham}, the Poisson
bivectors $L_{\tau_1}(P)$ and $L_{\tau_2}(P)$ are compatible if
and only if
$[L_{\tau_1}(L_{\tau_{2}}(P)),P]=0$.
\end{cor}
{\em Proof.} We have the following identity:
\be\label{t12}
0=L_{\tau_i}([L_{\tau_{3-i}}(P),P])
=[L_{\tau_i}(L_{\tau_{3-i}}(P)),P]+
[L_{\tau_i}(P),L_{\tau_{3-i}}(P)],\quad i=1,2.
\ee
Equations (\ref{t12}) for $i=1$ and $i=2$ are equivalent.
Indeed, subtracting (\ref{t12}) with $i=1$ from (\ref{t12}) with
$i=2$, and using the symmetry property (\ref{sym}) of the Schouten
bracket and the equality
$L_{\tau_1}L_{\tau_2}-L_{\tau_1}L_{\tau_2}=L_{[\tau_1,\tau_2]}$,
where $[\cdot,\cdot ]$ stands for the commutator of vector fields,
we obtain $[L_{[\tau_1,\tau_{2}]}(P),P]=0$. But this holds
automatically by virtue of (\ref{comp3}). Thus,
$[L_{\tau_1}(P),L_{\tau_{2}}(P)]=0$ if and only if
$[L_{\tau_1}(L_{\tau_{2}}(P)),P]=0$. $\square$

If $[\tau_1,\tau_{2}]=0$, $L_{\tau_1}(L_{\tau_{2}}(P))=0$, and
$L_{\tau_i}^2 (P)=0$, $i=1,2$, we recover Proposition 4.2 of
Smirnov \cite{smi}.\looseness=-1

For nondegenerate $P$ it is always possible
\cite{oevth} to find locally (and even globally, if the symplectic
form associated with $P^{-1}$ is exact) a `scaling' vector field $Z$
such that $L_Z(P)=P$. Then we can replace $\tilde\tau$ by
$\tilde\tau+\alpha Z$ and assume without loss of generality that
$\alpha=0$ in (\ref{comp2}). As we shall see below,
if $\dim\ker P\leq 1$ and $P$
is of locally constant rank, then the construction of
Corollary~\ref{inf_st} with $\alpha=0$ locally (and globally too,
if the second de Rham cohomology of $M$ is trivial and $P$ is
nondegenerate) yields {\em all} Poisson bivectors compatible with
$P$. \looseness=-1

Note that (\ref{comp2}) with $\alpha=0$ often holds for the
Poisson bivectors associated with
integrable systems, provided $\tau$ and $\tilde\tau$ are `first'
and `second' master symmetries for the latter, cf.\ e.g.\
\cite{dor, bl, oevth, oev, ffms}. Moreover, in presence of
a scaling symmetry we can
easily construct
the corresponding
$\tau$ and $\tilde\tau$
using the results of Oevel \cite{oevth, oev}.
Namely, the following assertions hold.
\looseness=-1
\begin{prop}\label{cpb}
Assume that $P$ and $\tilde P$ are Poisson bivectors, $P$ is
nondegenerate, and there exists a `scaling' vector field $\tau_0$
such that $L_{\tau_0} (P)=\mu P$ and $L_{\tau_0} (\tilde P)=\nu
\tilde P$ for some constants $\mu$ and $\nu$,
$\mu\neq\nu/2$.
Then $P$ and $\tilde P$ are compatible if and only if
\be\label{compsc} L_{R\tau_0}(P)=(2\mu-\nu)\tilde P, \ee where $R=\tilde P
P^{-1}$.
\end{prop}
{\em Proof.} If $[P, \tilde P]=0$, then (\ref{compsc}) holds by Theorem 2
of Oevel \cite{oev}. On the other hand, if (\ref{compsc}) holds,
then $P$ and $\tilde P$ are compatible by (\ref{comp3})
with $\tau=(1/(2\mu-\nu))R\tau_0$, as $\tilde P=L_{\tau}(P)$.
$\square$
\looseness=-1
\begin{cor}\label{cpbcor}
Let $P$ be a nondegenerate Poisson bivector and $\tilde P$ be a
bivector on $M$,
and let there exist a `scaling' vector field $\tau_0$ such that
$L_{\tau_0} (P)=\mu P$, $L_{\tau_0} (\tilde P)=\nu \tilde P$, and
(\ref{compsc}) holds with $R=\tilde P P^{-1}$ for some constants
$\mu$ and $\nu$ such that $\mu\neq\nu/2$ and $\mu\neq 2\nu/3$.
Then $\tilde P$ is a Poisson bivector, which is automatically compatible
with $P$, if and only if \be\label{poiss}
L^2_{R\tau_0}(P)=\frac{2\mu-\nu}{3\mu-2\nu}L_{R^2\tau_0}(P). \ee
\end{cor}
{\em Proof.} If $\tilde P$ is a Poisson bivector, then it is
compatible with $P$ by virtue of (\ref{compsc}) and (\ref{comp3}),
and (\ref{poiss}) 
holds by Theorem 2 of Oevel \cite{oev}.
On the other hand, let (\ref{poiss}) hold. By virtue of
(\ref{compsc}) we have $\tilde P=L_{\tau}(P)$ for
$\tau=R\tau_0/(2\mu-\nu)$, so (\ref{poiss}) implies that
(\ref{comp2}) holds for
$\tilde\tau=R^2\tau_0/((2\mu-\nu)(3\mu-2\nu))$, and $\tilde P$ is Poisson by
Corollary~\ref{inf_st}.  $\square$ \looseness=-1

Let e.g.\ $M=\mathbb{R}^{2m}$ with (global)
coordinates $x^1, \dots,\allowbreak x^{2m}$, and
$$
P_r=\left(\begin{array}{@{}rr@{}} 0 & \Lambda_r\\ -\Lambda_r & 0
\end{array}\right),
$$ where
$\Lambda_r=\mathop{\rm diag}\nolimits ((x^1)^r,\dots,(x^m)^r)$ is
a diagonal $m\times m$ matrix. Set
$\tau_0=(x^1,\dots,x^{m},0,\dots,0)^{T}$. We have
$L_{\tau_0}(P_r)=(r-1)P_r$. As $P_0$ obviously is a Poisson
bivector and
(\ref{compsc}) and (\ref{poiss}) hold for $P=P_0$, $\tilde P=P_r$,
$\mu=-1$ and $\nu=r-1$, by Corollary~\ref{cpbcor} for any $r$
$P_r$ is a Poisson bivector compatible with $P_0$, and
$P_r=L_{\tau_r}(P_0)$, where $\tau_r=(-(x^{1})^{r+1}/(r+1),\dots,
-(x^{m})^{r+1}/(r+1),0,\dots,0)^{T}$. Clearly,
$[L_{\tau_r}(L_{\tau_s}(P_0)),P_0]=\nobreak 0$,
so the Poisson bivectors $P_r$ and $P_s$ are compatible for any
$r$ and $s$ by Corollary~\ref{2t}. \looseness=-1

\section{Compatibility and Lichnerowicz--Poisson cohomology}
%
%
The condition $[L_{\tau}^2(P),P]=0$
is intimately related to the so-called Lichnerowicz--Poisson
cohomology introduced in \cite{li}. Indeed, the second
Lichnerowicz--Poisson cohomology $H^2_P(M)$ of a Poisson
manifold $(M,P)$ is precisely the set of bivectors $B$ solving
$[B,P]=0$ modulo the solutions of the form $B=L_{Y}(P)$, where $Y$
is a vector field on $M$.
Hence,
$[L_{\tau}^2(P),P]=0$
if and only if
there exist a vector field $\tilde\tau$ on $M$ and constants $a_i$
such that
\be\label{lpcoh}
L_{\tau}^2(P)=L_{\tilde\tau}(P)+\sum_{i=1}^{d}a_i B_i,
\ee
where
$B_i$, $i=1,\dots,d\equiv\dim H^2_P(M)$, form a basis of
$H^2_P(M)$. Thus, if $H^2_P(M)$ is known (see e.g.\ \cite{vai}
for a survey of results on its computation), Eq. (\ref{lpcoh})
yields a simple way to verify whether $L_{\tau}(P)$ is Poisson.

For instance, if $P$ is nondegenerate, $H^2_P(M)$ is
isomorphic \cite{li} to the second de Rham cohomology $H^2(M)$ of
$M$, so the following assertion holds. \looseness=-1
\begin{prop}\label{cpb2vf}
Suppose that $H^2(M)=0$, and let $P$ be a nondegenerate Poisson
bivector on $M$. Then a bivector $\tilde P$ on $M$ is a Poisson
bivector compatible with $P$ if and only if there exist vector
fields $\tau$ and $\tilde\tau$ on $M$ such that $\tilde
P=L_{\tau}(P)$ and $L_{\tau}^2 (P)= L_{\tilde\tau}(P)$.
\looseness=-1
\end{prop}
%
{\em Proof.}
For nondegenerate $P$ the condition $[B,P]=0$ is equivalent, cf.\
e.g.\ \cite{mor, li}, to $d(P^{-1}BP^{-1})=0$.
As $H^2(M)=0$, this is equivalent to $P^{-1}BP^{-1}=d\gamma_B$ for some
one-form $\gamma_B\in\wedge^1 M$. Upon setting $Y_B=-P\gamma_B$ we have
$B=L_{Y_B}(P)$, see e.g.\ \cite{oevth, mor}. Setting $\tau=Y_{\tilde
P}$ and $\tilde\tau=Y_{K}$, where $K=L_{\tau}^2(P)$, and using
Proposition~\ref{biham} completes the proof. $\square$\looseness=-1

As the second de Rham cohomology 
is always locally trivial, Proposition~\ref{cpb2vf}
{\em locally} describes {\em all} Poisson bivectors compatible
with a nondegenerate $P$ even if $H^2(M)\neq 0$. \looseness=-1
Moreover, if $\dim\ker P\leq 1$ on the whole of $M$ and $P$ is
of locally constant rank, then
$H^2_P(M)$ is locally trivial \cite{li},
i.e., for any bivector $B$ satisfying $[B,P]=\nolinebreak 0$ there still exists
a local (but not necessarily global) vector field $Y_B$ such that
$B=L_{Y_B}(P)$, so we arrive at the following result.
\looseness=-1
\begin{prop}\label{cpb2vfloc}
Let $P$ be a Poisson bivector of locally constant rank such that
$\dim\ker P\leq 1$ everywhere on $M$.
Then a bivector $\tilde P$ on $M$ is a Poisson
bivector compatible with $P$ if and only if there exist local
vector fields $\tau$ and $\tilde\tau$ such that
the equalities
$L_{\tau}^2 (P)=L_{\tilde\tau}(P)$ and
$\tilde P=L_{\tau}(P)$ hold locally. \looseness=-1
\end{prop}

By passing from the system of equations $[L_{\tau}^2
(P), P]=0$ for $\tau$ to $L_{\tau}^2 (P)=L_{\tilde\tau}(P)$ we
have essentially partially integrated the former, as $L_{\tau}^2
(P)=L_{\tilde\tau}(P)$ is a second order system of differential
equations with respect to $\tau$, while $[L_{\tau}^2 (P), P]=0$ is
of third order. Thus, solving $L_{\tau}^2 (P)=L_{\tilde\tau}(P)$
instead of $[L_{\tau}^2 (P), P]=0$ can considerably simplify the
search for and the classification of Poisson bivectors compatible
with $P$.
\looseness=-1
\section{Construction of integrable
bi-Hamiltonian dynamical systems}\label{bihamsec}
A vector field
$X$ on $M$ is called {\em locally Hamiltonian} with respect to a
Poisson bivector $P$, if $L_X(P)=0$, and {\em globally
Hamiltonian} w.r.t.~$P$, if it is locally Hamiltonian w.r.t.~$P$
and there exists a (smooth) function $H$ on $M$ such that $X=Pd
H$, cf.\ e.g.\ \cite{dor, bl, smi} and references therein.

Likewise, a vector field $X$ on $M$ is called {\em locally
bi-Hamiltonian} with respect to a pair of compatible Poisson
bivectors $P$ and $\tilde P$,
if $L_X(P)=L_{X}(\tilde P)=0$ on $M$, cf.\ e.g.\
\cite{magri, mamo, bl}. Finally, $X$ is
{\em globally bi-Hamiltonian}, if it is locally bi-Hamiltonian and
there exist smooth functions $H$ and $\tilde H$ on $M$ such that
$X=Pd H=\tilde Pd\tilde H$.
\looseness=-1
\begin{cor}\label{int}
Consider a Poisson bivector $P$ and a vector field $X$ on $M$ such
that $L_X(P)=0$. Assume that there exists a
vector field $\tau$ on $M$ such that $L_{\tau}(P)\neq 0$,
$L_{X}(L_{\tau}(P))=0$, and $[L_{\tau}^{2}(P),P]=0$. Then $X$ is
locally bi-Hamiltonian with respect to $P$ and $\tilde
P\equiv L_{\tau}(P)$.
If there also exist globally defined smooth functions $H$ and
$\tilde H$ on $M$ such that $X=Pd H=\tilde P d\tilde H$, then $X$
is globally bi-Hamiltonian on $M$. \looseness=-1
\end{cor}

This result
generalizes Theorem~5.1 of Smirnov
\cite{smi} and Corollary 1 of Chavchanidze \cite{chav}.
\looseness=-1

For instance, if we take $P$ from (\ref{blrwst}) and
$\tau=(-x-x^3/3, 0, z x^2-x^2 y^2)^{T}$, as in Section \ref{whenpb},
then $X=Pd H$ with $H=x^2 y^2-z x^2-z$ meets the requirements of
Corollary~\ref{int} and is globally bi-Hamiltonian:
$X=L_{\tau}(P)d\tilde H$ with $\tilde H=z$, cf.\ \cite{bl, blrw}.

We can invoke Corollaries~\ref{inf_st}
or \ref{cpbcor} or Propositions~\ref{cpb2vf} or \ref{cpb2vfloc}
in order to verify the condition $[L_{\tau}^{2}(P),P]=0$.
If $H^2(M)=0$ and $P$ is nondegenerate, then by
virtue of Proposition~\ref{cpb2vf}
the conditions of Corollary~\ref{int} are not
just sufficient but also {\em necessary} for a Hamiltonian w.r.t.
$P$ vector field $X$ to be bi-Hamiltonian, assuming that $P$ is
one of two compatible Poisson bivectors. Moreover, in this case
Proposition~\ref{cpb2vf} enables us to replace the condition
$[L_{\tau}^2(P),P]=0$ by the requirement of existence of a vector
field $\tilde\tau$ on $M$ such that $L_{\tau}^2(P)=L_{\tilde\tau}(P)$.
\looseness=-1

Likewise, Proposition~\ref{cpb2vfloc}
yields the following result.
\begin{cor}\label{intloc}
Consider a Poisson bivector $P$ of locally constant rank and a
vector field $X$ on $M$ such that $L_X(P)=0$ and $\dim\ker
P\leq\nobreak 1$ everywhere on $M$. Then $X$ is locally
bi-Hamiltonian with respect to $P$ and some other Poisson bivector
$\tilde P$ if and only if there exist local vector fields $\tau$
and $\tilde\tau$ such that locally we have $L_{\tau}(P)\neq 0$,
$L_{X}(L_{\tau}(P))=0$, $L_{\tau}^2(P)=L_{\tilde\tau}(P)$, and
$\tilde P$ can be locally written as $L_{\tau}(P)$.
\end{cor}

Under the assumptions of Corollary~\ref{int} or \ref{intloc}, suppose that at
least one of the Poisson bivectors $P$ and $\tilde P$ is
nondegenerate, so the dimension of $M$ is even.
Denote the nondegenerate bivector by $P_0$, and let $P_1$ stand
for the remaining Poisson bivector.
Then \cite{mamo, ff} $R=P_1 P_{0}^{-1}$ is a hereditary
recursion operator for the dynamical system $\dot x=X$, and
the eigenvalues of $R$ provide involutive integrals
for $\dot x=X$ \cite{mamo}.
In particular, if $R$ has a maximal possible number (i.e.,
$(1/2)\dim M$) of distinct eigenvalues, and all of them are
functionally independent, then
these eigenvalues form \cite{mamo} a complete set of functionally
independent involutive integrals for the dynamical system $\dot
x=X$, ensuring its complete integrability in the sense of
Liouville's theorem. \looseness=-1

Note that Corollaries~\ref{int} and \ref{intloc} readily
generalize to the case of {\em quasi-bi-Hamiltonian systems}
considered e.g.\ in \cite{bl, qbh}, when
we have $L_{\rho X}(\tilde P)=0$ for some
smooth function $\rho$ on $M$ instead of $L_X(\tilde P)=0$: it suffices to replace the
conditions $L_X(L_{\tau}(P))=0$ and $X=\tilde P d\tilde H$
by $L_{\rho X}(L_{\tau}(P))=0$ and
$X=(1/\rho)\tilde P d\tilde H$ respectively.
\looseness=-1

\medskip

\noindent{\bf Remark 1}
Recall that given two compatible Poisson bivectors
$P_0$ and $P_1$, a Poisson pencil $\mathcal{P}$ associated
with $P_0$ and $P_1$ is the set of all linear combinations
of the form $\lambda P_0+\mu P_1$, where $\lambda$
and $\mu$ are constants. The compatibility of $P_0$ and $P_1$
implies that any such linear combination again is a Poisson bivector.
A vector field $X$ is said to be locally bi-Hamiltonian
with respect to the Poisson pencil associated
with $P_0$ and $P_1$ if $L_X(\lambda P_0+\mu P_1)=0$ for any $\lambda$ and $\mu$.

Let us call a Poisson pencil $\mathcal{P}$ {\em regular} (resp.\ {\em weakly regular})
if there exist $\lambda$ and $\mu$ such that
$P=\lambda P_0+\mu P_1$ is nondegenerate (resp. $P$ is of locally constant rank
on $M$ and $\dim\ker P\leq\nobreak 1$ everywhere on $M$).
Let $\tilde P=\tilde \lambda P_0+\tilde \mu P_1$ be linearly independent of $P$.
Then Proposition~\ref{cpb2vf} yields the following result:

Suppose that $H^2(M)=0$. Then a vector field $X$ on $M$ is locally
bi-Hamiltonian with respect to a regular Poisson pencil $\mathcal{P}$
if and only if $L_X(P)=0$ and there exist vector fields $\tau$
and $\tilde\tau$ such that $L_{\tau}(P)\neq 0$,
$L_{X}(L_{\tau}(P))=0$, $L_{\tau}^2(P)=L_{\tilde\tau}(P)$, and
$\tilde P$ can be written as $L_{\tau}(P)$.
\looseness=-1

Likewise, Corollary~\ref{intloc} is equivalent to the following assertion:

A vector field $X$ on $M$ is locally
bi-Hamiltonian with respect to a weakly regular Poisson pencil $\mathcal{P}$
if and only if $L_X(P)=0$ and there exist local vector fields $\tau$
and $\tilde\tau$ such that locally we have $L_{\tau}(P)\neq 0$,
$L_{X}(L_{\tau}(P))=0$, $L_{\tau}^2(P)=L_{\tilde\tau}(P)$, and
$\tilde P$ can be locally written as $L_{\tau}(P)$.
\looseness=-1

Thus, we have obtained necessary and sufficient conditions for existence of
a (weakly) regular Poisson pencil such that
a given vector field $X$ is locally bi-Hamiltonian with respect to this pencil.
\looseness=-1
\section{Infinite-dimensional case}\label{infdim}
Up to now we assumed that $M$ is a finite-dimensional manifold.
Nevertheless, all the above results, except for
Proposition~\ref{cpb2vfloc}, Corollary~\ref{intloc}  and the
statement on Liouville integrability of $\dot x=X$,
readily extend to the case of Hamiltonian formalism
over the complex over a Lie algebra $\mathfrak{A}$ associated with
a representation $\pi$ of $\mathfrak{A}$,
see Example 2.2 of \cite{dor}. This setting is very general and
naturally includes the most of interesting cases when the
underlying manifold $M$ is infinite-dimensional. \looseness=-1

The desired extension is achieved by passing from
bivectors to the skew-symmetric operators
$P:\Omega^1\rightarrow\mathfrak{A}$,
where $\Omega^1$ is the set of all linear mappings from
$\mathfrak{A}$ to $\pi$, and replacing a)~the notion of
nondegeneracy of a bivector by that of invertibility of the
operator, b)~the condition $H^2(M)=0$ by the requirement of triviality
of the second cohomology for the complex in question.
The standard Hamiltonian formalism over a finite-dimensional
manifold $M$ is recovered if we take for the
complex in question the de Rham complex of $M$ \cite{dor}. \looseness=-1

Moreover, Propositions~\ref{biham} and \ref{cpb}
and Corollaries~\ref{inf_st}--\ref{int} in fact remain valid
(after performing the above replacement) within the framework of Hamiltonian
formalism over an arbitrary $(\Omega,d)$-complex over a Lie
algebra $\mathfrak{A}$ with nondegenerate pairing between
$\mathfrak{A}$ and $\Omega^1$, see Ch.~2 of~\cite{dor} for more
details on such complexes.

The key example of an infinite-dimensional
($\Omega,d$)-comp\-lex undoubtedly
is that of formal calculus of variations, see e.g.\ \cite{dor}
for further details, and cf.\ \cite{kr} for
a somewhat different approach to the
Hamiltonian formalism for PDEs.
Let us briefly recall some basic
properties of this complex for
the case of one space variable $x$ ($x\in\mathbb{R}$ or $x\in
S^1$) and $n$ dependent variables, essentially following \cite{dor, olv_eng2}.
\looseness=-1

Consider an algebra $\mathcal{A}_{j}$ of locally analytic
functions of $x,t,\bu,\bu_{1},\dots,\allowbreak \bu_{j}$, where
$\bu_{k}=(u_{k}^{1},\dots,u_{k}^{n})^{T}$ are $n$-component vectors,
$\bi{u}_0\equiv\bi{u}$, and let
$\mathcal{A}=\bigcup_{j=0}^{\infty}\mathcal{A}_{j}$.  We
shall call the elements of $\mathcal{A}$ {\em local} functions.
Let us make $\mathcal{A}$ into a differential algebra by
introducing a derivation
$$
D\equiv D_x=\p/\p x+\sum\limits_{j=0}^{\infty}\bu_{j+1}\p/\p\bu_j.
$$
Denote by $\Im D$
the image of $D$ in $\mathcal{A}$, and let
$\tilde\mathcal{A}=\mathcal{A}/\Im D$. Following the tradition,
denote the canonical projection $\rho:\mathcal{A}\rightarrow
\tilde\mathcal{A}$ by $\int dx$.
Then we have
the following `formal integration by parts' formula valid for any
$f,g\in\mathcal{A}$:
$$
\int f D(g) dx=-\int g D(f) dx.
$$
Informally, $x$ plays the role of the space variable, and $D$ is
the total $x$-derivative, cf.\ e.g.\ \cite{olv_eng2}, so one often
writes $\p^j \bi{u}/\p x^j$ instead of $\bi{u}_j$.

We take for $\mathfrak{A}$ the Lie algebra of evolution vector
fields (EVFs) of the form
$X\!=\!\sum_{m=0}^{\infty}\!\sum_{p=1}^{n}\! D^m(h^p)\p/\p
u^{p}_m$;
$\bi{h}=(h^1,\dots,h^n)^T$ is called the cha\-racteristics of $X$,
$h^p\in\mathcal{A}$.
The characteristics of the commutator $[X,Y]$ of two EVFs is given
by $Y(\bi{h})-X(\bi{g})$, where $\bi{h}$ and $\bi{g}$ are
characteristics of $X$ and $Y$, respectively, see e.g.\
\cite{olv_eng2}. Clearly, the characteristics are in one-to-one
correspondence with the EVFs, so in what follows we shall identify
the EVFs with their characteristics, cf.\ e.g.\ \cite{olv_eng2}.
\looseness=-1

We have $\Omega^0=\tilde\mathcal{A}$, and  $\Omega^1$ consists of
the vertical one-forms $\omega= \sum_{p=1}^{n} \gamma_p
d u^{p}$, where $\gamma_p\in\mathcal{A}$. The pairing
between $\mathfrak{A}$ and $\Omega^1$ is given by
$$
\langle X,\omega \rangle=\int \sum_{p=1}^{n} \gamma_p h^p dx.
$$

Introduce in $\mathcal{A}$ the operator of variational
derivative (see e.g.\ \cite{dor, olv_eng2})
$$
\delta/\delta\bu=\sum\limits_{m=0}^{\infty}(-D)^{m}\p/\p\bu_m.
$$
Then the differential of $F=\int f dx\in\tilde\mathcal{A}$ reads
$d F=\sum_{r=1}^{n}\delta f/\delta u^r du^r$.



Let $P$
be a Hamiltonian operator of the form
$$
P=\sum\limits_{m=0}^{q}a_{m}D^{m}+\sum
\limits_{\kappa=1}^{p}\bi{G}_{\kappa}\otimes D^{-1}\circ
\boldsymbol{\gamma}_{\kappa},
$$
where $q\geq 0$, $a_i$ are $s\times s$ matrices with entries from
$\mathcal{A}$,
$\bi{G}_{\kappa},\boldsymbol{\gamma}_{\kappa}\in\mathcal{A}^s$
(in fact, $\bi{G}_{\kappa},\boldsymbol{\gamma}_{\kappa}\in\mathfrak{A}$).
Then, or even more broadly, for $P$ that can be written as a
formal series of the form $\sum_{m=-\infty}^{q}a_{m}D^{m}$,
Propositions~\ref{biham},~\ref{cpb} and~\ref{cpb2vfloc} and
Corollaries \ref{inf_st}--\ref{intloc} remain valid, if we replace
the requirement of nondegeneracy of $P$ or the condition of local
constancy of rank of $P$ along with $\dim\ker P\leq 1$ by
$\det a_q \neq 0$ and allow for nonlocalities
(like e.g.\ $D^{-1}(\zeta)$ for some $\zeta\in\mathcal{A}$) in the
characteristics of $\tau$ and $\tilde\tau$.
\looseness=-1

Consider e.g.\ the operators of the form
$$
\begin{array}{l}
\hspace*{-0.7ex}P^{ij}=g^{ij}(\bi{u})D
+\!\!\sum\limits_{k=1}^{n}\! b^{ij}_{k}(\bi{u})u_x^{k}
+\!\!\sum\limits_{\alpha=1}^{r}
\sum\limits_{k,l=1}^{n}\!\!\epsilon_{\alpha}
(w^{\alpha}(\bi{u}))^{i}_{k}u_x^{k} D^{-1}\circ
(w^{\alpha}(\bi{u}))^{j}_{l}u_x^{l}\label{ferbv1},\\
\hspace*{-0.7ex}\tilde P^{ij}=\tilde g^{ij}(\bi{u})D
+\!\!\sum\limits_{k=1}^{n}\!\tilde b^{ij}_{k}(\bi{u})u_x^{k}
+\!\!\sum\limits_{\alpha=1}^{\tilde r}\sum\limits_{k,l=1}^{n}
\!\!\tilde\epsilon_{\alpha} (\tilde w^{\alpha}(\bi{u}))^{i}_{k}u_x^{k} D^{-1}\circ
(\tilde w^{\alpha}(\bi{u}))^{j}_{l}u_x^{l}\label{ferbv2},
\end{array}
$$
where $\epsilon_{\alpha}$ and
$\tilde\epsilon_{\alpha}$ are constants satisfying
$(\epsilon_{\alpha})^2=1$ and $(\tilde\epsilon_{\alpha})^2=1$.

These operators
have a common scaling $\tau_0=x\bi{u}_x$. Hence, if $\det g^{ij}\neq 0$,
Proposition~\ref{cpb} and Corollary~\ref{cpbcor}
with $\tau_0=x\bi{u}_x$ and $\mu=\nu=1$
provide easily verifiable criteria
for compatibility of $P$ and $\tilde P$ and
for $\tilde P=L_{\tau}(P)$ to be a Hamiltonian operator if so is $P$.
The operators of this type were introduced by Ferapontov \cite{fer2},
and we refer the reader to this paper
for the conditions under which these operators are Hamiltonian
and the discussion of their properties.
\looseness=-1


Consider now the complex of formal calculus of variations
for the case
of two
dependent variables $u^1\equiv u$
and $u^2\equiv v$ (so $u_x\equiv u^1_1$,
$v_x\equiv u^2_1$), and two
skew-symmetric operators \cite{kup}
$$
P=\left(\begin{array}{@{}rr@{}} 0 & D\\ D & 2D
\end{array}\right),
\tilde P=\left(\begin{array}{@{}cc@{}} 2 u D+u_x & -D^2+v D+2 u
D+2u_x\\ D^2+v D+v_x +2 u D & 4 v D+2 v_x
\end{array}\right).
$$
Informally,
the role of manifold $M$ is played here by an appropriate
functional space (e.g.\ the Schwartz space) of two-component smooth
functions $(u,v)$ of $x$, cf.\ e.g.\ \cite{dor, bl}. \looseness=-1

For $\tau_0=(x u_x+u, xv_x+v)^T$ we have
$L_{\tau_0}(P)=0$ and $L_{\tau_0}(\tilde P)=\tilde P$.
The operator $P$ is obviously
Hamiltonian, as it is of odd order and has constant coefficients,
cf.\ e.g.\ \cite{dor, olv_eng2}.
Since (\ref{compsc}) and (\ref{poiss}) are easily seen to hold
with $\mu=0$ and $\nu=1$, our Corollary~\ref{cpbcor} reconfirms
that $\tilde P$ is \cite{kup} a Hamiltonian operator compatible
with $P$. \looseness=-1

What is more \cite{kup}, $X=(\frac{1}{2}D(-u_x+2u v-u^2),
\frac{1}{2}D(v_x-2 u_x- 2 u^2+2 u v +v^2))^T$ is Hamiltonian with
respect to $P$: $X=Pd H$ for $H=(1/2)\int (-u_x v- u^2 v+ u v^2)dx
$. We also have $\tilde P=L_{\tau}(P)$, where $\tau=-\tilde
P\gamma$, $\gamma=(-2 x u+ xv, xu)^T$, and $L_X(\tilde P)=0$, so
by Corollary~\ref{int} $X$ is locally bi-Hamiltonian with respect
to $P$ and $\tilde P=L_{\tau}(P)$. In fact, $X$ is globally
bi-Hamiltonian \cite{kup}, as $X=\tilde Pd\tilde H$ for $\tilde
H=(1/2) \int (u v-v^2)dx$.
\looseness=-1

Thus, we have reconfirmed (cf.\ \cite{kup})
the bi-Hamiltonian nature of
the modified dispersive water wave
system
$$ u_t=\frac{1}{2}D(-u_x+2u v-u^2),\quad
v_t=\frac{1}{2} D(v_x-2
u_x- 2 u^2+2 u v +v^2). $$

For another example, consider the complex of formal calculus of
variations~\cite{dor} for the case of one space variable $x$ and
$n$ dependent variables $u^i$, and a skew-symmetric operator $P$
of Dubrovin--Novikov
type \cite{dn, dn84}, cf.\ also \cite{fer2},
\be\label{dn0} {P}^{ij}=g^{ij}(\bi{u})D +
\sum_{k=1}^{n}b^{ij}_{k}(\bi{u})u_x^{k}, \ee where
$\bi{u}=(u^1,\dots,u^n)^{T}$,
$u^k_x\equiv u^k_1$,
and the indices $i,j,k,\dots$ run from $1$ to $n$.
The role of the manifold $M$ is now played by the loop space,
i.e., the space of smooth mappings from $S^1$ with a local
coordinate $x$ to an $n$-dimensional manifold $N$ with local
coordinates $u^i$, see e.g.\ \cite{m1,dn,dn84} for details.
\looseness=-1

Recall \cite{dn, dn84} that $P$ (\ref{dn0}) with $\det g^{ij}\neq 0$
is a Hamiltonian operator if and only if $g^{ij}$ is a flat
(pseudo-\nobreak) Riemannian metric on $N$ and $b^{ij}_{k}=
-\sum\nolimits_{m=1}^{n}g^{im}\Gamma^{j}_{mk}$, where
$\Gamma^{j}_{mk}$ is the Levi-Civita connection associated with
$g^{ij}$: $\Gamma^{k}_{ij}=(1/2)\sum_{s=1}^{n}g^{ks}
(\p g^{sj}/\p x^i+\p g^{is}/\p x^j-\p g^{ij}/\p x^s)$.

For $P$ (\ref{dn0}) and $\tau=\tau(\bi{u})$ we have
(see e.g.\ \cite{m1})
$$
\ba{l}
(L_{\tau}(P))^{ij}=\sum\limits_{s=1}^{n}
\left( \tau^s  {\p g^{ij}\over\p u^s} - g^{sj}
{\p\tau^i\over\p u^s} - g^{is} {\p \tau^j\over\p u^s} \right)D
+ \sum\limits_{s,k=1}^{n}\left(\tau^s {\p b^{ij}_k\over\p u^s}
-b^{sj}_k {\p \tau^i \over \p u^s} - b^{is}_k {\p \tau^j\over\p u^s}
+ b^{ij}_s {\p \tau^s\over\p u^k} -
g^{is} {\p^2 \tau^j\over\p u^s \p u^k } \right) u^k_x.
\ea
$$
\begin{prop}\label{cpbdn} Let
$P$ be a Hamiltonian operator of the form (\ref{dn0}) with $\det
g^{ij}\neq \nolinebreak0$.
Then a skew-symmetric operator $\tilde P$ of Dubrovin--Novikov
type is a Hamiltonian operator compatible with $P$ if and only if
there exist local vector fields $\tau=\tau(\bi{u})$
and $\tilde\tau=\tilde\tau(\bi{u})$
on $N$ such that
locally we have $\tilde P=L_{\tau}(P)$ and  $L_{\tau}^2(P)
=L_{\tilde\tau}(P)$. \looseness=-1
\end{prop}
Let us stress that $\tau$ and $\tilde\tau$ depend here solely on
$\bi{u}$. In particular, they do not involve $\bi{u}_j$ with $j\geq 1$.
Proposition \ref{cpbdn} is proved
along the very same lines as
Proposition~\ref{cpb2vfloc},
with usage of the following lemma.
\begin{lem}\label{cpbdnlem} Suppose that
$P$ is a Hamiltonian operator of the form (\ref{dn0}) with $\det
g^{ij}\neq 0$.
Then for any skew-symmetric operator $\tilde P$ of
Dubrovin--Novikov type (not necessarily Hamiltonian) that
satisfies $[\tilde P,P]=0$ there exists a local vector field
$\tau=\tau(\bi{u})$ on $N$
such that $\tilde P$ can be written locally as $\tilde
P=L_{\tau}(P)$.
\end{lem}

{\em Proof of the lemma.}
According to
\cite{dn, dn84} (see also \cite{getz} for a more formal setting) any Hamiltonian
operator~$P$ of the form (\ref{dn0}) with $\det g^{ij}\neq 0$ can
be locally transformed into
a Hamiltonian operator 
with constant coefficients of the form
\be\label{canon} P_{\rm can}^{ij}=g_0^{ij}D,
\ee
where $g_0^{ij}=0$ for $i\neq j$ and
$g_0^{ii}$ satisfy $(g_0^{ii})^2=1$, via an invertible transformation
$\bi{u}\mapsto\tilde\bi{u}=\bi{f}(\bi{u})$.
But for $P=P_{\rm can}$ our lemma holds
by Proposition 1 of \cite{magri_hd}, cf.\ also \cite{m1}, and we
obtain the desired result in full generality
by just 
going back from $\tilde\bi{u}$ to $\bi{u}$.
$\square$ \looseness=-1

Thus, the classification of compatible Hamiltonian operators of
Dub\-ro\-vin--Novikov type essentially amounts, at least locally, to
the classification of pairs of vector fields $\tau$ and
$\tilde\tau$ on $N$ such that $L_{\tau}^2(P) =L_{\tilde\tau}(P)$
for $P=P_{\rm can}$ (\ref{canon}).
In particular, in this way we can recover
the quasihomogeneous Hamiltonian pairs constructed by
Dubrovin in \cite{dubr} using the theory of Frobenius manifolds.
For $\tilde\tau=0$ we
come back to the case analysed by Fordy and Mokhov \cite{fm}.
The comparison of the classification
results obtained in our approach with e.g.\ those of Mokhov \cite{mo02}
will be the subject of our future work.

Consider now the systems of hydrodynamic type, that is, the
systems of the form $\bi{u}_t=A(\bi{u})\bi{u}_x$, where
$A(\bi{u})$ is an $n\times n$ matrix \cite{dn,dn84}.
Proposition~\ref{cpbdn} immediately yields the following analog of
Corollary~\ref{intloc}.

\begin{cor}\label{intlocdn}
Consider a Poisson bivector $P$ of Dubrovin--Novikov type with
$\det g^{ij}\neq 0$ and
an evolutionary vector field $X=(A(\bi{u})\bi{u}_x)\p/\p\bi{u}$
such that $L_X(P)=0$. Then $X$ is locally bi-Hamiltonian with
respect to $P$ and some other Poisson bivector $\tilde P$ of
Dubrovin--Novikov type if and only if there exist local vector
fields $\tau(\bi{u})$ and $\tilde\tau(\bi{u})$ on $N$ such that
locally we have $L_{\tau}(P)\neq 0$, $L_{X}(L_{\tau}(P))=0$,
$L_{\tau}^2(P)=L_{\tilde\tau}(P)$, and $\tilde P$ can be locally
written as $L_{\tau}(P)$.
\looseness=-1
\end{cor}
%


\noindent{\bf Remark 2}
In analogy with Remark 1, let us call a Poisson pencil $\mathcal{P}$
associated with compatible Hamiltonian operators $P_0,P_1$ of Dubrovin--Novikov type {\em regular}
if there exist constants $\lambda$ and $\mu$ such that $P=\lambda P_0+\mu P_1$
has $\det g^{ij}\neq 0$. Let $\tilde P=\tilde \lambda P_0+\tilde \mu P_1$ be linearly independent of $P$.
Then we can restate Corollary~\ref{intlocdn} 
as follows:
\looseness=-1

An evolutionary vector field $X=(A(\bi{u})\bi{u}_x)\p/\p\bi{u}$ is locally
bi-Hamiltonian with respect to a regular Poisson pencil $\mathcal{P}$
of Hamiltonian operators of Dubrovin--Novikov type if and only
if $L_X(P)=0$ and there exist local vector fields $\tau$
and $\tilde\tau$ on $N$ such that locally we have $L_{\tau}(P)\neq 0$,
$L_{X}(L_{\tau}(P))=0$, $L_{\tau}^2(P)=L_{\tilde\tau}(P)$, and
$\tilde P$ can be locally written as $L_{\tau}(P)$.
\looseness=-1

This reformulation of Corollary~\ref{intlocdn}
gives necessary and sufficient conditions for
existence of a regular Poisson pencil
of Hamiltonian operators of Dubrovin--Novikov type
such that a given EVF $(A(\bi{u})\bi{u}_x)\p/\p\bi{u}$
is locally bi-Hamiltonian with respect to this Poisson pencil (cf.\ Remark~1).
Moreover, we expect that this result
can be efficiently employed for the
classification of such EVFs and associated bi-Hamiltonian hydrodynamic type systems,
and we plan to address this issue elsewhere.

\section{Conclusions}
We have shown above that
$[L_{\tau}^2(P),P]=0$ is a
necessary and sufficient condition for $\tilde P=L_{\tau}(P)$ to
be a Poisson bivector, if so is $P$. This enabled us to benefit
from the powerful results of \cite{li} on the
Lichnerowicz--Poisson cohomology and
obtain new easily verifiable necessary and sufficient conditions for
existence of Poisson bivectors compatible with a given Poisson
bivector and for transformability of a given 
dynamical system
into a bi-Hamiltonian form. \looseness=-1

For instance, when $P$ and $\tilde P$
have a common scaling, the verification
of their compatibility and of the conditions for
$\tilde P =L_{\tau}(P)$ to be a Poisson bivector
along the lines of Proposition~\ref{cpb} and
Corollary~\ref{cpbcor} is considerably easier than the
cumbersome direct computation
of the corresponding Schouten brackets,
especially in the infinite-dimensional case,
as
the application of our results
requires just the computation of Lie derivatives
that can be readily performed using
the 
computer algebra software like {\it Maple} or {\it
Mathematica}. \looseness=-1

Moreover, unlike e.g.\ Theorem 5.1 of \cite{smi},
our results 
provide not
merely sufficient but
{\em necessary and sufficient} conditions
ensuring that for a given dynamical system $\dot x=X$
there exists a regular or weakly regular Poisson pencil
such that $\dot x=X$ is locally
bi-Hamiltonian with respect to this pencil,
see Remark~1 for details.
We believe that
this result will enable one
to perform a complete
classification of
such dynamical systems. 
\looseness=-1

The results of Propositions~\ref{cpb2vf},
\ref{cpb2vfloc} and \ref{cpbdn} enable us to perform, at least in
certain cases, a `partial integration' of the system
$[\tilde P,P]=[\tilde P,\tilde P]=0$ by
replacing it with $\tilde P=L_{\tau}(P)$ and
$L_{\tau}^2(P)=L_{\tilde\tau}(P)$. The latter system is much
easier to solve just because
it is generally easier to find a
couple of vectors than a skew-symmetric rank two tensor,
cf.\
\cite{smi}.
\looseness=-1

Thus, we can classify the Poisson bivectors compatible with a
given Poisson bivector $P$ of locally constant rank  with
$\dim\ker P\leq 1$ using Propositions~\ref{cpb2vf}
and~\ref{cpb2vfloc} after bringing $P$ into canonical form using
the Darboux theorem. We expect that a similar approach, based on
the local description of $H^2_P(M)$, could be extended to the case
of $\dim\ker P>1$ as well. The classification of Hamiltonian
operators of Dubrovin--Novikov type also can be performed in a
similar fashion, see Section~\ref{infdim} above. It would be
interesting to compare the results obtained in this way with those
found by other known classification methods, for instance from
\cite{gz}--\cite{fer}, and \cite{mo02}, and to find out whether
one can extend the results of present paper to the case of
Dirac~\cite{dor} and Jacobi (see e.g.~\cite{jac}) structures. We
plan to address these issues in our future work. \looseness=-2


\subsection*{Acknowledgements}

I am sincerely grateful to Profs.\ M.~B\l aszak, L.~Hlavat\'y,
P.J. Olver, V.V.~Sokolov and Drs.\
E.V.~Ferapontov,
M.V.~Pavlov, M. Pedroni and
R.G.~Smirnov for stimulating discussions. I am also pleased
to thank Profs.\ M.~B\l aszak, C.~Morosi, W.~Oe\-vel, F.~Turiel
and Drs.\ M.~Pedroni, F.~Petalidou and R.G.~Smirnov for
kindly providing me with the copies of \cite{bl},
\cite{mamo, mor},
\cite{oevth}, \cite{tur}, \cite{cas},
\cite{peth}, \cite{oev, smi}, respectively.
Last but not least, I express deep
gratitude to the organizers of $11^{th}$~Colloquium ``Quantum
Groups and Integrable Systems" (June 20-22, 2002, Prague,
Czech Republic) where most of the results of this paper were
presented, for their kind hospitality. \looseness=-1

This research was supported in part by DFG via Graduiertenkolleg
``Geometrie und Nichtlineare Analysis'' at Institut f\"ur
Mathematik of Humboldt-Universit\"at zu Berlin, where the author
held a postdoctoral fellowship in 2001, and by the Jacob Blaustein
postdoctoral fellowship.
I also acknowledge with gratitude
the support from the Czech Grant Agency under
grant No.~201/00/0724 and the
Ministry of Education, Youth and Sports of Czech Republic under
grant MSM:J10/98:192400002. \looseness=-2




\begin{thebibliography}{99}
\bibitem{magri}Magri, F.:
A simple model of the integrable Hamiltonian equation,
{\em J. Math. Phys.} {\bf 19} (1978), 1156--1162.

\bibitem{gd}Gelfand, I.M., Dorfman, I.Ya.:
Hamiltonian operators and algebraic structures related to
them,
{\it Funct. Anal. Appl.} {\bf 13} (1979), 248--262.

\bibitem{mamo}Magri, F. and Morosi, C.:
A geometrical characterization of integrable Hamiltonian systems
through the theory of Poisson--Nijenhuis manifolds,
{\it Quaderno} {\bf S~19} (1984),
Dip. Matematica, Universit\`a degli Studi di Milano, Milan, 1984.

\bibitem{ff}Fuchssteiner, B., and Fokas, A.S.:
Symplectic structures, their B\"acklund transformations and
hereditary symmetries, {\it Physica D} {\bf 4} (1981/82), no. 1,
47--66.
\looseness=-1

\bibitem{dor}Dorfman, I.:
{\it Dirac Structures and Integrability of
Nonlinear Evolution Equations},
John Wiley \& Sons, Chichester, 1993.

\bibitem{olv_eng2}
Olver, P.J.: {\it Applications of Lie Groups to Differential
Equations}, Springer, New York, 1993.

\bibitem{bl}
B\l aszak, M.:
{\it Multi-Hamiltonian Theory of Dynamical
Systems},
Springer, Heidelberg, 1998.


\bibitem{gz}Gelfand, I.M. and Zakharevich, I.:
On the local geometry of a bi-Hamiltonian structure,
in: L. Corwin (ed), {\it The Gelfand Seminars, 1990-1992},
Birkh\"auser, Basel, pp.~51--112.

\bibitem{tur}Turiel, F.:
Classification locale simultan\'ee de deux formes
symplectiques compatibles,
{\it Manuscripta Math.} {\bf 82} (1994), no. 3-4, 349--362.
%

\bibitem{tur2}Turiel, F.:
Tissus de Veronese analytiques de codimension sup\'erieure et
structures bihamiltoniennes,
{\it C. R. Acad. Sci. Paris} S\'er. I Math. {\bf 331} (1) (2000),
no. 1, 61--64.


\bibitem{fer} Ferapontov, E.V.:
Compatible Poisson brackets of hydrodynamic type,
{\it J. Phys. A: Math. Gen.} {\bf 34} (2001), 2377--2388, arXiV: {\tt math.DG/0005221}


\bibitem{fm} Fordy, A.P., Mokhov O.I.:
On a special class of compatible Poisson structures of hydrodynamic type,
{\it Physica D} {\bf 152/153} (2001), 475--490.

\bibitem{m1}Mokhov, O.I.: Compatible
Dubrovin--Novikov Hamiltonian operators, Lie derivative and
integrable systems of hydrodynamic type, to appear in {\it
Nonlinear evolution equations and dynamical systems (Cambridge,
July 24-30, 2001)}
({\it Preprint} {\tt math.DG/0201281})

\bibitem{mks} Kosmann-Schwarzbach, Y., Magri, F.:
Lax-Nijenhuis operators for integrable systems.
{\it J. Math. Phys.} {\bf 37} (1996), no. 12,
6173--6197.

\bibitem{olv-bih} Olver, P.J.:
Canonical forms and integrability of bi-Hamiltonian systems,
{\it Phys. Lett. A} {\bf 148} (1990), no. 3-4,
177--187.

\bibitem{fern}Fernandes, R.L.:
Completely integrable bi-Hamiltonian systems,
{\it J. Dynam.\ Differential Equations} {\bf 6} (1994), no. 1,
53--69.

\bibitem{oevth}
Oevel, W.: {\it Rekursionmechanismen f\"ur Symmetrien und
Erhaltungss\"atze in Integrablen Systemen}, Ph.D. thesis,
University of Paderborn, Paderborn, 1984. \looseness=-1

\bibitem{oev}
Oevel, W.:
A geometrical approach to integrable
systems admitting time dependent invariants,
in M. Ablowitz, B. Fuchssteiner, M. Kruskal (eds)
{\it Topics in Soliton Theory and Exactly Solvable Nonlinear
Equations (Oberwolfach, 1986)}
World Scientific Publishing, Singapore, 1987,
pp.~108--124. \looseness=-1

\bibitem{mor}Morosi, C.:
Some remarks on the construction
of bi-Hamiltonian manifolds.
{\it Atti Sem. Mat. Fis. Univ. Modena} {\bf 36}
(1988), no. 2, 303--320.

\bibitem{mma}Magnano, G., Magri, F.:
Poisson--Nijenhuis structures and Sato hierarchy,
{\it Rev. Math. Phys.} {\bf 3} (1991), no. 4,
403--466.

\bibitem{magri2} Magri, F.:
On the geometry of soliton equations,
{\it Acta Appl. Math.} {\bf 41} (1995), 247--270.
\looseness=-1

\bibitem{cas} Casati, P., Falqui, G., Magri, F., Pedroni, M.,
The KP Theory revisited II. The reduction theory and the
Adler--Gel'fand--Dickey brackets, Preprint
SISSA/3/96/FM,
Trieste 1996.

\bibitem{fal}Falqui, G., Magri, F., Pedroni, M.,
Bi-Hamiltonian geometry, Darboux coverings, and
linearization of the KP hierarchy, {\it Comm. Math. Phys.}
{\bf 197} (1998), 
303--324, arXiv: {\tt solv-int/9806002}

\bibitem{smi} Smirnov, R.G.:
Bi-Hamiltonian formalism: a constructive approach,
{\it Lett. Math. Phys.} {\bf 41} (1997), 333--347.

\bibitem{peth} Petalidou, F.: {\it \' Etude
locale de structures bihamiltoniennes}, Th\`ese de Doctorat,
Universit\'e Pierre et Marie Curie, Paris, 1998.

\bibitem{pet}Petalidou, F.:
Sur la symplectisation de
structures bihamiltoniennes,
{\it Bull. Sci. Math.} {\bf 124}(2000), no. 4,
255--286.

\bibitem{li}Lichnerowicz, A.:
Les vari\'et\'es de Poisson et leurs alg\`ebres
de Lie associ\'ees,
{\it J. Differential Geometry} {\bf 12} (1977),
no. 2, 253--300.

\bibitem{ffms} Fokas, A.S., Fuchssteiner, B.:
The hierarchy of the Benjamin-Ono equation.
{\it Phys. Lett. A} {\bf 86}(1981), no. 6-7,
341--345.

\bibitem{bih-prepr}Sergyeyev, A.: A simple way to make a
Hamiltonian system into bi-Hamiltonian one, Preprint GA~4/2002,
Silesian University in Opava, Czech Republic, May 2002.

\bibitem{chav}Chavchanidze, G.: Non-Noether symmetries and their
influence on phase space geometry, {\it J. Geom. Phys.} {\bf 48}
(2003),
190--202, arXiV: {\tt math-ph/0211014}

\bibitem{dn}
Dubrovin, B.A., Novikov, S.P.:
Hamiltonian formalism of one-dimensional systems of the
hydrodynamic type and the
Bogolyubov-Whitham averaging method,
{\it Soviet Math. Dokl.} {\bf 27} (1983),
665--669.

\bibitem{dn84}
Dubrovin, B.A., Novikov, S.P.:
On Poisson brackets of hydrodynamic type,
{\it Soviet Math. Dokl.} {\bf 30} (1984), 651--654.



\bibitem{bl1}B\l aszak, M.:
Degenerate Poisson pencils on curves: new separability theory,
{\it J. Nonlinear Math. Phys.} {\bf 7} (2000), no. 2,
213--243, arXiv: {\tt math.DG/9803106}


\bibitem{bl2}B\l aszak, M.:
From bi-Hamiltonian geometry to separation of variables:
stationary Harry-Dym and the KdV dressing chain,
{\it J. Nonlinear Math. Phys.} {\bf 9} (2002), Supplement, no. 2,
1--13, arXiv: {\tt nlin.SI/0202001}


\bibitem{blrw} B\l aszak, M., and Wojciechowski, S.:
Bi-Hamiltonian dynamical systems related to low-dimensional Lie algebras,
{\it Physica A} {\bf 155} (1989), no. 3,
545--564.
%
\bibitem{vai}Vaisman, I.: {\it Lectures on the geometry of Poisson manifolds},
Birkh\"auser, Basel, 1994.


\bibitem{qbh}
Tondo, G. and Morosi, C.:
Bi-Hamiltonian manifolds, quasi-bi-Hamiltonian systems and
separation variables,
{\it Rep. Math. Phys.} {\bf 44} (1999), no. 1-2,
255--266, arXiv: {\tt solv-int/9811008}

\bibitem{kr} Krasil'shchik, I. S.:
Cohomological approach to Poisson structures
on nonlinear evolution equations,
{\it Lobachevskii J. Math.} {\bf 3} (1999), 127--145.

\bibitem{fer2} Ferapontov, E.V.:
Differential geometry of nonlocal Hamiltonian operators of hydrodynamic type,
{\it Funct. Anal. Appl.} {\bf 25} (1991), 195--204.




\bibitem{kup}Kupershmidt, B.A.:
Mathematics of dispersive water waves,
{\it Comm. Math. Phys.} {\bf 99} (1985), no. 1,
51--73.



\bibitem{getz}Getzler, E.:
A Darboux theorem for Hamiltonian operators in the formal calculus of variations.
{\it Duke Math. J.} {\bf 111} (2002), no. 3, 535--560,
arXiv: {\tt math.DG/0002164}


\bibitem{magri_hd}Degiovanni, L., Magri, F., Sciacca, V.:
{\it On deformation of Poisson manifolds of hydrodynamic type},
Preprint {\tt nlin.SI/0103052} (2001)

\bibitem{dubr}Dubrovin, B.A.:
Flat pencils of metrics and Frobenius manifolds, in:
{\em Integrable systems and algebraic geometry},
Saito M.-H. (ed.) et al.,
World Scientific, Singapore, 1998, pp. 47--72,
arXiv: {\tt math.DG/9803106}

\bibitem{mo02}Mokhov, O.I.:
On the integrability of equations
for nonsingular pairs of compatible flat metrics,
{\it Theoret. and Math. Phys.} {\bf 130} (2002), no. 2, 198--212.


\bibitem{jac} Guedira, F., Lichnerowicz, A.:
G\'eom\'etrie des alg\`ebres de Lie locales de Kirillov,
{\it J. Math. Pures Appl.} {\bf 63} (1984), no. 4,
407--484.

\end{thebibliography}
\end{document}